\documentclass[12pt]{iopart}
\usepackage{graphicx}
\usepackage{color}
\usepackage[utf8]{inputenc}
\usepackage{amsfonts}
\usepackage{amssymb}
\usepackage{iopams} 

\usepackage[normalem]{ulem}

\begin{document}
\title{Half-metallic ferromagnetism in layered CdOHCl induced by hole doping}
\author{H Banerjee$^1$*, P Barone$^2$, 
	S Picozzi$^2$}

\address{$^1$ The Abdus Salam Internatinal Centre for Theoretical Physics (ICTP), Str. Costiera, 11, 34151 Trieste TS, Italy, Currently at Institute of Theoretical and Computational Physics, Graz University of Technology, NAWI Graz, Petersga{\ss}e 16, Graz, 8010, Austria.} 
\address{$^2$ Consiglio Nazionale Delle Ricerche (CNR) SPIN, Universita degli studi G. d'Annunzio di Chieti, via dei Vestini, 31, 66100, Chieti, Italy.}
\ead{h.banerjee10@gmail.com}

\begin{indented}
\item[August 2020]
\end{indented}

\begin{abstract}

Next-generation spintronic devices will benefit from low-dimensionality, ferromagnetism, and half--metallicity, possibly controlled by electric fields. We find these technologically--appealing features to be combined with an exotic microscopic origin of magnetism in doped CdOHCl, a van der Waals material from which 2D layers may be exfoliated. 
By means of first principles simulations, we 
predict  homogeneous hole--doping  to give rise to $p$-band magnetism in both the bulk and monolayer phases and interpret our findings in terms of Stoner instability: as the Fermi level is tuned via hole--doping through  singularities in the 2D-like density of states, ferromagnetism develops with large saturation magnetization of 1 $\mu_B$ per hole, leading to a half-metallic behaviour for layer carrier densities of the order of 10$^{14}$ cm$^{-2}$. Furthermore, we put forward electrostatic doping as an additional handle to induce magnetism in monolayers and bilayers of CdOHCl. Upon application of critical electric fields perpendicular to atomically--thin-films (as low as 0.2 V/\AA$\:$ and 0.5 V/\AA$\:$ in the bilayer and monolayer case, respectively), we envisage the emergence of a magnetic half-metallic state. The different  behaviour of monolayer vs bilayer systems, as well as an observed asymmetric response to positive and negative electric fields in bilayers, are interpreted in terms of intrinsic polarity of  CdOHCl atomic stacks, a distinctive feature of the material. In perspective, given the experimentally accessible magnitude of critical fields in bilayer of CdOHCl, one can envisage $p$ band magnetism to be exploited in miniaturized spintronic devices. 


\end{abstract}
%
\vspace{2pc}

\section{Introduction}
Since isolated graphene sheets have been successfully exfoliated more than ten years ago, low-dimensional materials have been object of intense research activity aiming at tailoring and optimizing their properties for next-generation electronic and optoelectronic applications. Several two-dimensional (2D) materials complementary to graphene have been since identified, synthesized and intensively studied, including graphene-like 2D-Xenes (X=Si,Ge,Sn),
transition-metal mono- and dichalcogenides (e.g., GaX and MoX$_2$, with X=S, Se, Te) or even layered oxides as $\alpha$-SnO or ZnO\cite{2dmat_synthesis2020}. 
Many more potential 2D materials have been proposed within the family of existing van der Waals (vdW) layered materials \cite{Mounet2018}, leading to a database of exfoliable compounds whose detailed analysis could significantly contribute to further progress in the field. 
The potential impact of such low-dimensional materials is two-fold. On the one hand, 
their reduced dimensionality meets the technological requirements for continuous device miniaturization. On the other hand, the physical properties often change dramatically as the number of layers is reduced, not only providing a new degree of freedom for potential applications but also resulting in novel and exotic physical phenomena, triggered, e.g., by enhanced correlation effects 
and electronic instabilities.
%
For instance, the possibility of inducing and controlling magnetism in  2D crystals has been extensively studied, in view of potential spintronic applications that could possibly push the scaling limits and/or the power consumption beyond those of charge-based nanodevices\cite{spintronic_review2020}. First-principles calculations have predicted that local magnetic moment can be induced by nonmagnetic chemical dopants or vacancies in  graphene, 2D boron nitride, ZnO, and MoS$_2$\cite{liu2007,oleg2007,guo2012,zhou2013}. More recently, an intense research activity has been devoted to the emergence of spontaneous ferromagnetism without doping in 2D materials, after its observation in monolayers and few-layers of CrI$_3$ and CrGeTe$_3$ \cite{Gong2017, Huang2017}. The inherent magnetism in these systems originates from partially occupied $d$ shells of the transition-metal atoms, while the long-range ferromagnetic phase is allowed by the relatively strong magnetic anisotropy due to spin-orbit coupling\cite{Klein1218, Jiang2018, Torelli_2018, Xu2018}. Because of their small-gap semiconducting character, such 2D ferromagnets can be used as spin filters or magnetic tunnel junctions\cite{spintronic_rev2019}. Nevertheless, the efficiency of such devices would greatly benefit if ferromagnetism develops together with half metallicity, where electrons with one spin orientation display a metallic nature coexisting with the insulating character of opposite-spin ones\cite{degroot_prl1983,pickett_2001,Hirohata_spintronicrev_2014}. Because of the ideal 100\% spin-polarization at the Fermi level, half-metallic ferromagnets are also ideal candidates for spin-injection applications\cite{Hirohata_spintronicrev_2014,ramsteiner_hm_spininj_prb2008}.

Half-metallic magnetism has been indeed proposed 
in the semiconducting monochalcogenide GaSe monolayer\cite{Louie}, where a Stoner-type itinerant magnetic instability has been predicted to be triggered by a high density of states (DOS) close to the Fermi level. In the monolayer limit, the band structure of GaSe around the valence band maximum develops an inverted Mexican-hat-like energy dispersion, thus giving rise to a van Hove singularity with power-law divergent DOS close to the Fermi energy. By tuning the Fermi level via hole doping through the van Hove singularity, Stoner ferromagnetism can be induced as $N(\epsilon_F)\,I > 1$, where $N(\epsilon_F)$ is the density of states at the Fermi energy and $I$ is the Stoner parameter measuring the exchange energy. By means of first-principles calculations, an averaged magnetic moment of 1$\mu_B$/hole was predicted for a range of hole doping with carrier densities of 10$^{13}$-10$^{14}$/cm$^2$ \cite{Louie} and an estimated mean-field transition temperature of 90~$K$\cite{Louie}.
The same mechanism has been theoretically predicted in other 2D materials displaying Mexican-hat-induced van Hove singularities, such as InSe\cite{InSe_acsanm.8b01476}, phosphorene and arsenene\cite{Fu_2dmater_2017}, InP$_3$ \cite{jacs2017}, PtS$_2$\cite{konstantina} and $\alpha$-SnO\cite{Castro-neto}. Indeed, as high-order van Hove singularities  lead to power-law divergent DOS in the 2D limit\cite{Fu_ncomm2019}, a Stoner-like instability has also been predicted in other 2D materials upon carrier doping. Among these, we mention 
PdSe$_2$\cite{Bang-Gui}, with an estimated transition temperature reaching 800~$K$, $\alpha/\beta-$In$_2$Se$_3$\cite{Li2019}, proposed as active element for spin-polarized and bipolar field-effect spin-filter devices, and two-dimensional holey C$_2$N\cite{Gong_C2N_2017}, where electron doping was proposed to drive the half-metallic ferromagnetic transition. 

Most of the theoretical predictions were based on first-principles calculations where the carrier density was tuned by
changing the total number of electrons in the unit cell and adding a compensating jellium background of opposite charge
to preserve charge neutrality in the simulations. A practical, viable path towards achieving the desired carrier density could be the conventional chemical doping, i.e., the introduction of $p$-type defects and dopants, as extensively done in semiconductor-based devices. However, besides the technological limitations of such approach in nanometer-scale systems (where, e.g., it is extremely challenging to control the distribution of dopants and to tailor their concentration), it is often found that substitutional doping results in spin-polarization densities mainly localized on defect/impurity locations\cite{InSe_acsanm.8b01476,Li2019,konstantina,ZHAO20161}, at the expenses of the system's transport properties. On the other hand, gate-controlled electrostatic doping has rapidly emerged as a very effective technique to control and manipulate carrier densities in atomically thin materials\cite{gupta_review2017}, allowing to achieve high carrier densities exceeding 10$^{14}$ cm$^{-2}$, e.g., in graphene\cite{graphene_gating} and MoS$_2$\cite{mos2_gating}. Using an electrolyte as a gate dielectric of a field-effect transitor with a 1 nm-thick  electric  double  layer  geometry, electric fields as high as 0.5 V/\AA~ can be attained \cite{Ueno_apl2010}. It would be therefore highly desirable to identify a suitable 2D material where half-metallic ferromagnetism can be induced and tuned by gate controlled hole-doping.



In our study we focus on a simple van der Waals semiconductor, CdOHCl\cite{CdOHCl_stru}, whose 3D structure was taken from the Crystallographic Open Database\cite{COD_2012}. In its bulk form, CdOHCl crystallizes in a layered structure with an underlying hexagonal lattice. The completely filled $d$-shell of the transition metal Cd prevents the stabilization of localized magnetic phases. Nevertheless, the quasi-2D nature of the electronic structure, reflecting the underlying layered crystal, may lead to high density of states and van Hove singularities close to the valence-band maxima. By means of ab-initio  simulations, we here explore the possibility of driving CdOHCl (both in the bulk as well as thinned down to monolayer or bilayer phases) to a magnetic and half--metallic ground state by means of homogeneous hole-doping as well as upon application of electric fields.

\section{Computational Details}
Our first-principles calculations were based on a plane wave basis set, as implemented in the Vienna Ab-initio Simulation Package (VASP) \cite{vasp} with projector-augmented wave (PAW) pseudopotentials\cite{paw}. The exchange-correlation functional was chosen according to the Perdew Burke Ernzerhof (PBE) \cite{pbe} implementation of the generalised gradient approximation (GGA). 
The homogeneous doping was  tuned by changing the total number of electrons in the unit cell and by adding
a compensating jellium background of opposite charge to preserve charge neutrality.
A large Monkhorst-Pack mesh was used with the maximum grid of $64\times64\times1$ for monolayer, 
$36\times36\times2$ for bilayer, and $12\times12\times4$ for bulk calculations, with a plane wave cutoff of 500 eV for all calculations. The tetrahedron method was  used for accurate integration of the DOS in the Brillouin zone. 
For the calculations carried out on  thin film geometries, we used a supercell approach, where the thickness of the vacuum layer was chosen to be at least 20 \AA. We remark that large thicknesses of vacuum ($\sim$ 35 \AA) were required for calculations with  CdOHCl bilayers, in order to achieve a proper convergence in the presence of electric fields. The effect of  spurious electric fields in the vacuum, due to periodic boundary conditions, was taken care of by means of dipole corrections, as implemented in VASP.  
The van der Waals corrections (\'a la DFT-D3 method of Grimme with zero damping) were also included in our VASP simulations. Lattice constants were kept fixed to experimental values, whereas internal atomic positions were relaxed until forces reached values below 0.005 eV/\AA.


\section{Bulk CdOHCl: Structural and electronic properties}

Bulk CdOHCl crystallizes in the  hexagonal lattice  (space group $P6_3mc$) shown in Figures 1 (a), (b), with lattice parameters  $a$=$b$=3.665~\AA, $c$=10.230~\AA. \cite{cdohcl-cryst-struct}
 Figure 1 (a) shows  the layered structure of CdOHCl, with  a single layer highlighted. It is evident that CdOHCl is a van der Waals (vdW) crystal, characterized by a large relative distance  ($\sim$3.2 \AA) between the layers; as such, the crystal is likely to be easily exfoliated. 
Indeed, we evaluated the exfoliation energy by comparing the energy difference per unit area between bulk (per layer) and an ``isolated" monolayer of CdOHCl. 
This method has been shown by Jung {\em et al.} to be equivalent to traditional methods to calculate exfoliation energies, however with a  lower computational effort \cite{Jung2018}. In our case, the exfoliation energy results in an experimentally accessible value of $\sim$ 36 meV/\AA$^2$, well below the threshold value used in \cite{Mounet2018} to classify ``easily exfoliable" van der Waals compounds.

We remark that each CdOHCl layer shows a Cl-Cd-O-H atomic stacking along the $z$ axis perpendicular to the layer and, therefore, two different terminations [H on one side and Cl on the other side, cfr Figure 1 (a)], resulting in a strong polarity along the $z$ axis. When layers are stacked in the bulk phase, the crystal structure preserves a polar axis (consistent with the $P6_3mc$ polar space group)  which will play an important role in what follows. 

\begin{figure}[h]
\centering
	\includegraphics[width=0.85\columnwidth]{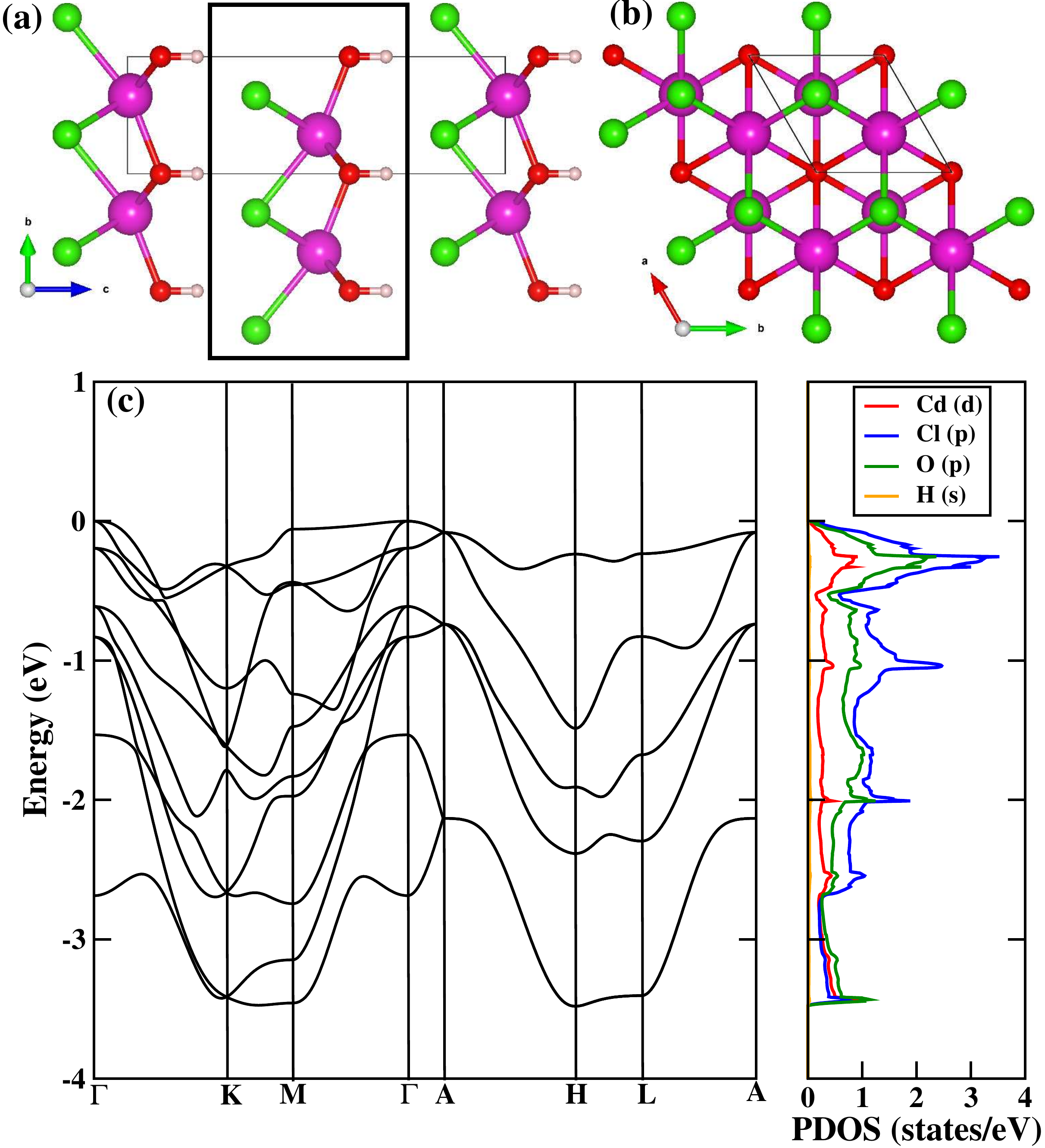}
	\caption{(Colour online)
The top panels show the crystal structure of bulk  CdOHCl. Magenta, green, red and light-pink spheres represent  Cd, Cl, O and H atoms, respectively.  Panel (a) shows the side view, showing the layered nature of the vdW crystal with a monolayer marked by the box. Panel (b) shows the top view, demonstrating the hexagonal symmetry of the crystal structure when viewed from the $z$ direction, for a single layer of CdOHCl. Panel (c) shows  the electronic structure of Bulk CdOHCl. The  left part shows the band structure, 
and the right part shows the  projected density of states, showing explicitly the contribution from different atoms.}
\end{figure}

The bulk electronic structure shows the ground state of CdOHCl to be insulating and non-magnetic, as seen from the Density of States (DOS) and band structure presented in Figure 1 (c). The insulating bands show an admixture of Cl and O $p$ bands with a large dominance of Cl bands near the Fermi energy. The $d$ bands of Cd lie deep in the valence bands (below -4 eV) and are completely filled; hence,  neither they play a relevant role in the electronic bonding of the system, nor they give rise to any magnetic feature. 
Notably, a very high and relatively broad feature is observed in the DOS around -0.25 eV with respect to the Fermi energy, corresponding to the extremal points at $H$ and $L$ and to the weakly dispersing band along the $H-L-A$ high-symmetry line. The presence of such high DOS not far from the Fermi energy suggests that the system is prone to electronic instabilities and, specifically, to Stoner-type ferromagnetic transition,  as will be discussed in the following section. 

\section{Hole doping: bulk vs monolayer CdOHCl}

We here investigate the effect of doping CdOHCl in its bulk and monolayer phases  and focus on the resulting electronic structure. 
The systems are homogeneously doped with fractional holes, uniformly distributing them all over the material (or, equivalently,  removing some fraction of electrons from the total electronic counting). We note that we also considered substitutional doping ({\em i.e.} by replacing Cl with S atom); however, since a controlled selective doping is likely difficult to be achieved in experiments and it theoretically leads to results similar  to those for the homogeneous doping, we don't discuss it further and concentrate on homogeneous doping  in what follows. 

\subsection{Doping of bulk structure}

\begin{figure}
    \includegraphics[width=\columnwidth]{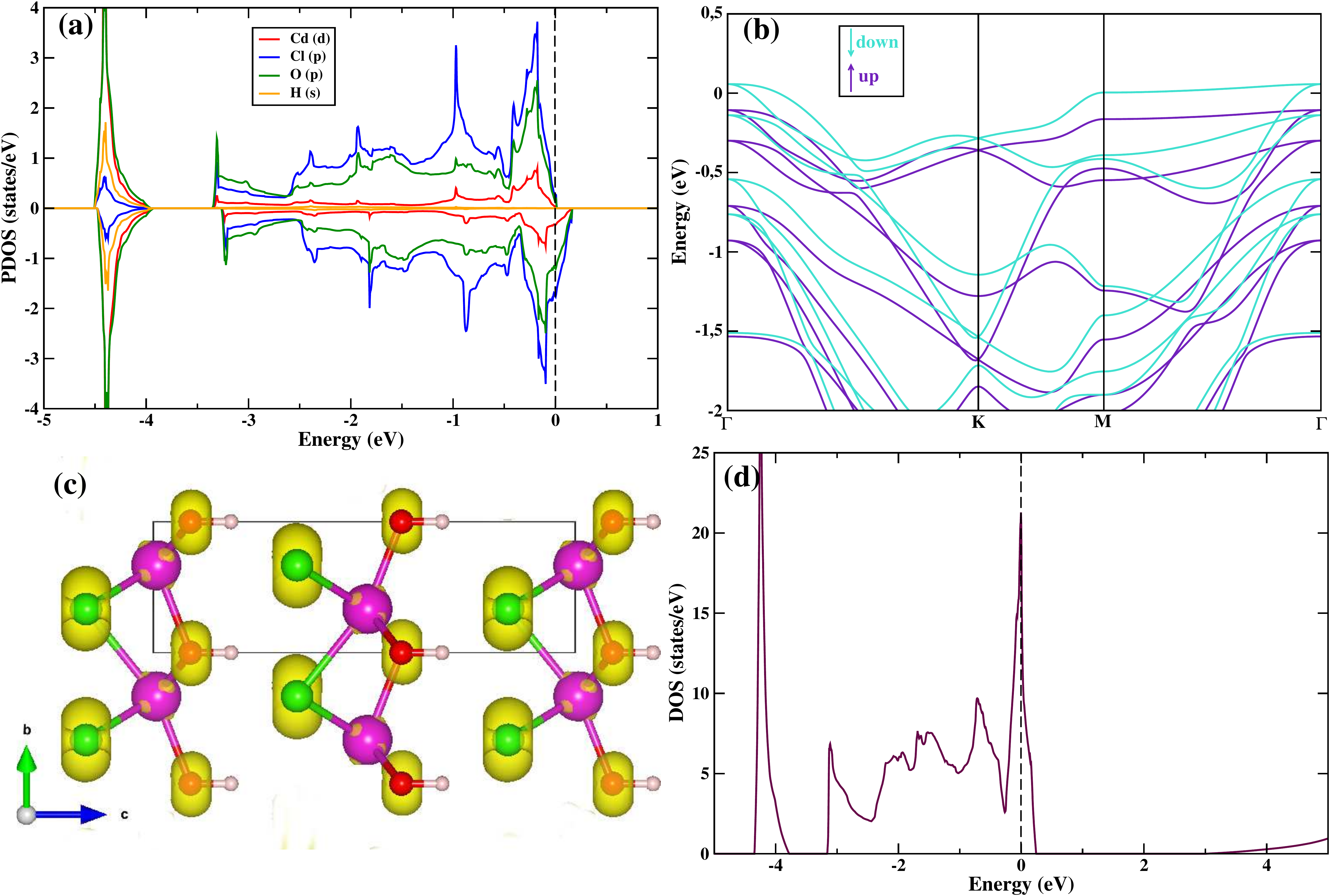}
	\caption{(Colour online)
\color{black} Electronic structure for 0.4 hole doping in bulk CdOHCl. Panel (a) shows the projected density of states projected on the different relevant orbitals. Panel (b) shows the spin resolved band structure with the up (down) spin channel marked with indigo (torquoise) spin channel. Panel (c) shows the magnetisation density in the system (atomic color labelling same as in Figure 1). Panel (d) shows the total DOS in a paramagnetic case with 0.4 hole doping.}
\end{figure}

We carried out uniform doping of the bulk crystal with various fractional number of holes, ranging from 0.1 to 2 holes. Since in the unit cell there are two formula--units ({\em i.e.} two layers of CdOHCl), this amounts to a range from 0.05 holes/formula--unit(fu) upto 1 hole/fu. We observe that, upon doping, the system shows a spontaneous spin polarisation,  the minimum ``critical" hole doping required to make it magnetic being 0.2 holes per fu ({\em i.e.} 0.4 holes in the bulk). The electronic structure for this case is shown in Figure 2. As evident  in both the projected DOS (cfr Figure 2(a)) and the spin resolved band structure (cfr Figure 2(b)), we obtain a half metallic ground state, {\em i.e.} a metallic (insulating) behaviour of spin-down (spin-up) states.

The projected DOS shows that the states at the Fermi energy have a predominant contribution from Cl-$p$ ($\sim$0.075 $\mu_B$) states with some  small contribution from O-$p$ states. Cl -$p$ orbitals have higher moments compared to O-$p$ ($\sim$0.055 $\mu_B$) orbitals with little to no contribution from Cd-$d$ or H-$s$ orbitals to the magnetization, and the total magnetic moment of course being equal to the number of holes created. The arising of a $p$ orbital magnetism, distributed among Cl-$p$ and O-$p$ orbitals, is also confirmed by the spin density plotted in   Figure 2 (c).
We remark that higher doping fractions lead to a similar electronic structure, with  magnetic moments increasing on Cl-$p$ and O-$p$ orbitals upon increasing hole concentrations. 

Interestingly, a non--magnetic calculation at the critical hole doping reveals that the Fermi energy is tuned at a sharp peak in the DOS, as displayed in Figure 2 (d), a situation compatible with the Stoner mechanism proposed for GaSe and other 2D systems\cite{Louie, InSe_acsanm.8b01476,Fu_2dmater_2017,jacs2017,konstantina,Castro-neto,Bang-Gui,Li2019,Gong_C2N_2017}. It turns out that the DOS at the Fermi level, $N(\epsilon_F)$, exceeds 5 states/(eV spin f.u.); on the other hand, the Stoner parameter can be estimated as $I = \mu_B\Delta E/M$, where $\Delta E$ is the exchange splitting energy of two spin channels near $\epsilon_F$ and $M$ is the magnetic moment per formula unit. For 0.2 hole/f.u. we have $\Delta E = 0.17~$eV and $M=0.2 \mu_B$, resulting in $I=0.85~$eV. Even though it is well known that the Stoner parameter is usually overestimated by roughly $20\%$ within the local spin-density approximation\cite{Ortenzi2012}, the Stoner criterion $N(\epsilon_F)I>1$ is clearly fulfilled in hole-doped CdOHCl, confirming that the half-metallic ferromagnetic phase stems from the Stoner mechanism of itinerant magnetism.

\subsection{Doping of monolayer structure}
When moving to homogeneous doping of monolayers, the situation appears to be very similar to the bulk. The critical doping fraction is evaluated as 0.2 holes per layer, consistently with what already found for the bulk. 
As seen from Figure 3, upon including the critical doping fraction of holes, a half metallic ground state emerges, with corresponding spin--polarization predominantly in Cl $p$ and O $p$ bands. Here Cl $p$ and O $p$ orbitals show moments of $\sim$ 0.12 $\mu_B$ and 0.03 $\mu_B$ respectively. In Figure 3 (b) and (d) a comparison between bulk and monolayer electronic structures is reported, in the spin--polarized (Figure 3 b) and paramagnetic (Figure 3 d) state. In Figure 3 (b) the monolayer bands (solid lines) are compared with the bulk
bands (dashed lines); the two appear to be rather similar, expect for the fact that the number of bands is clearly higher in the bulk,  due to the higher number of fu in the bulk unit cell. In Figure 3(d) the paramagnetic DOS for both bulk (dashed) and monolayers (solid) are reported, showing the van Hove singularity peak basically at the same position in energy and corresponding to the Fermi level. 
As such, we propose that the observed $p$-band magnetism originates from
a Stoner mechanism with an exchange-field splitting of the
electronic states near the VBM in both the monolayer and bulk phase. A tentative mean-field estimate of the transition temperature $T_C$ can be obtained by minimizing the electronic free energy of the system at finite temperatures, where the temperature is tuned by changing the smearing factor ($\sigma = K_B T$) in the Fermi-Dirac distribution function. Such procedure allows to track the temperature dependence of the spontaneous magnetization and, finally, the Curie transition temperature\cite{Louie}. For monolayer CdOHCl with 0.2 hole doping, we find that the critical smearing factor is 0.055 eV, corresponding to $T_C$=638K, much higher than the Curie temperature predicted for GaSe\cite{Louie} but of the same order of magnitude of estimated $T_C$ for other analogous 2D half-metallic ferromagnets \cite{Bang-Gui, Fu_2dmater_2017}. 


\begin{figure}
  \includegraphics[width=\columnwidth]{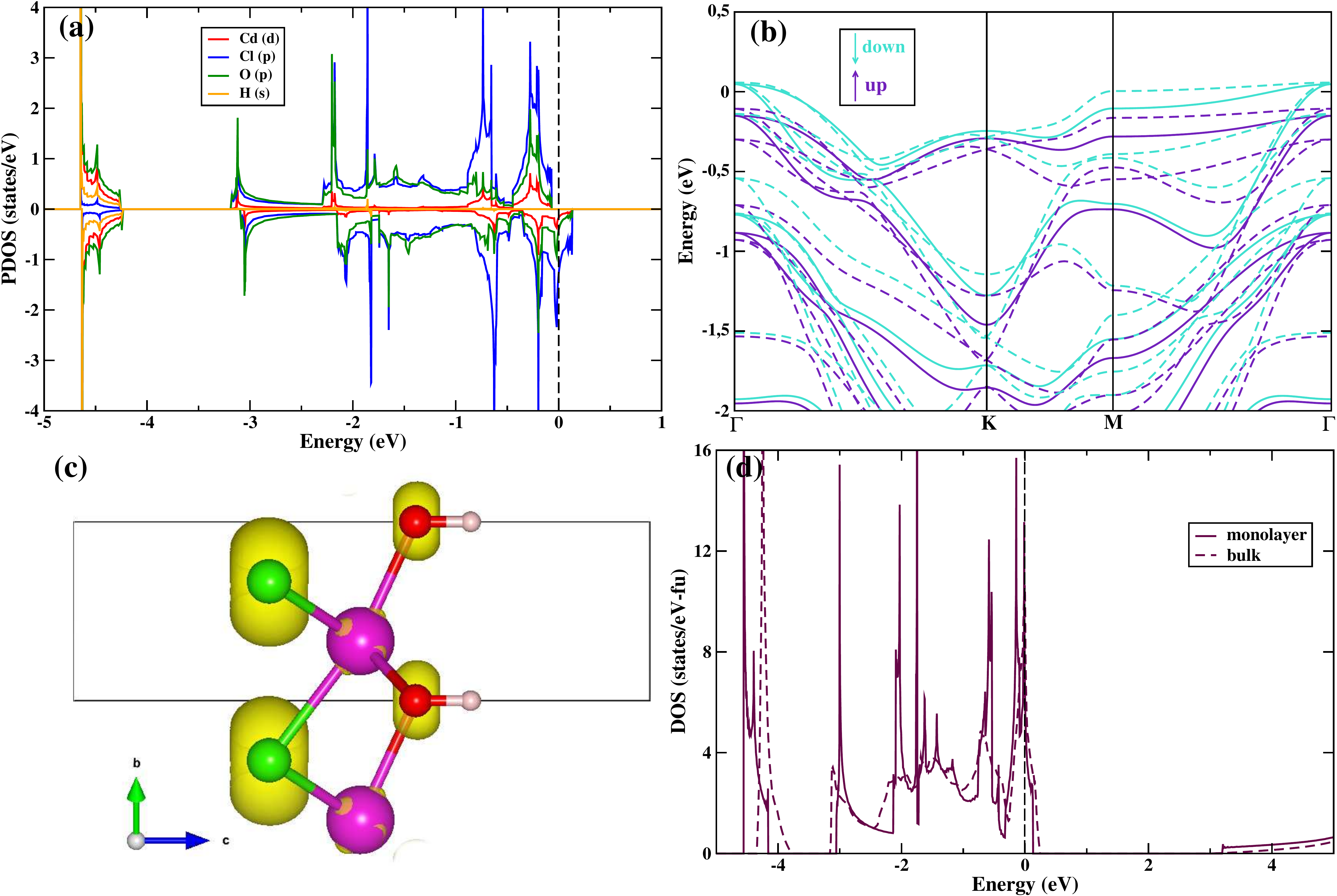}
	\caption 
	{(Colour online) 
\color{black} Electronic structure for the case of 0.2 hole doping in the monolayer of CdOHCl. Panel (a) shows the projected density of states projected to the different relevant orbitals of the atoms. Panel (b) shows the spin resolved band structure with the up (down) spin channel marked with indigo  (torquoise). 
The bulk bands are shown here with dashed lines for comparison (indigo and turquoise for up and down spins, respectively).  Panel (c) shows the magnetisation density in the monolayer (atomic colors follow the same scheme as in Figure 1). Panel (d) shows the total DOS in the paramagnetic case with 0.2 hole doping. 
The bulk paramagnetic DOS is also marked here with dashed line for comparison.}
\end{figure}

In table 1 we summarize some properties of interest in the case of hole doped CdOHCl. In closer detail, we report the critical induced magnetic moment, which for the monolayer reaches a value of 1$\mu_B$/hole, as also observed by Louie in GaSe monolayer {\em et al.} \cite{Louie}. Our critical carrier density of 1.78$\times$10$^{14}$/cm$^2$ is well within the range of 10$^{13}$-10$^{14}$/cm$^2$ for stable ferromagnetism observed by the same authors  and in the class of analogous 2D half-metallic ferromagnets\cite{InSe_acsanm.8b01476, Fu_2dmater_2017,jacs2017,konstantina, Castro-neto, Bang-Gui, Li2019, Gong_C2N_2017}. 

\begin{table}[]
    \centering
\begin{tabular}{|c|c|c|}
	\hline 
{\bf Property}	& {\bf Bulk} & {\bf Monolayer} \\ 
	\hline 
Critical induced magnetic moment ($\mu_B$/hole)	& 0.93  & 1.0 \\ 
	\hline 
Critical Carrier density/layer	& 1.77$\times$10$^{14}$/cm$^2$ & 1.78$\times$10$^{14}$/cm$^2$\\ 
	\hline 
Spin Polarisation Energy $E_{PM}-E_{FM}$ (meV/hole)	& 6.2  & 5.1 \\ 
	\hline 
\end{tabular} 

\caption{Relevant properties: comparison between bulk and monolayer.  The critical magnetic moment is defined as the magnetic moment at the threshold doping fraction which gives rise to stable magnetic moments, per unit doping fraction. The critical carrier density per layer is calculated as the doping fraction which gives rise to stable magnetic moment per unit area of the $ab$ lattice plane. The spin polarization energy is defined as the energy difference between the paramagnetic and the ferromagnetic phase at the critical carrier density per unit doping fraction.}
    \label{tab1}
\end{table}



\section{Effect of Electric field on CdOHCl films}
In view of an experimental route to gate--controlled electrostatic doping,  we focus in this section on the effects of an applied electric field on CdOHCl films. The emergence of a magnetization as a function of electric field can be rationalized as follows.
A constant electric field  (or, equivalently, a potential with constant slope as a function of the $z$ direction) can be considered as ``doping" the material by affecting its electronic structure ({\em i.e.} through selective shifting of different orbitals), finally driving it into a magnetic ground state. An example of this mechanism at play is indeed found for both monolayer and bilayer: both systems undergo a transition to a spin polarized state upon application of electric fields. However, the comparison shows that the electronic structure of monolayers and bilayers and their response to applied fields are quite different, as detailed below. 

\subsection{Electric field applied to monolayer}

Let us first discuss the monolayer case upon application of electric field. As can be seen from Figure 4 (a), for the specific case of an electric field of 0.6V/$\AA$,  a shift primarily in the Cl $p$ bands is observed. The effect of electrostatic doping is therefore maximum on the Cl atoms: a large magnetic moment is induced on the Cl ion, while almost no magnetization is present on the oxygen atoms. This behaviour is correspondingly seen in the plot of magnetisation density in  Figure 4 (d).

\begin{figure}[h]
	\centering
	\includegraphics[width=\columnwidth]{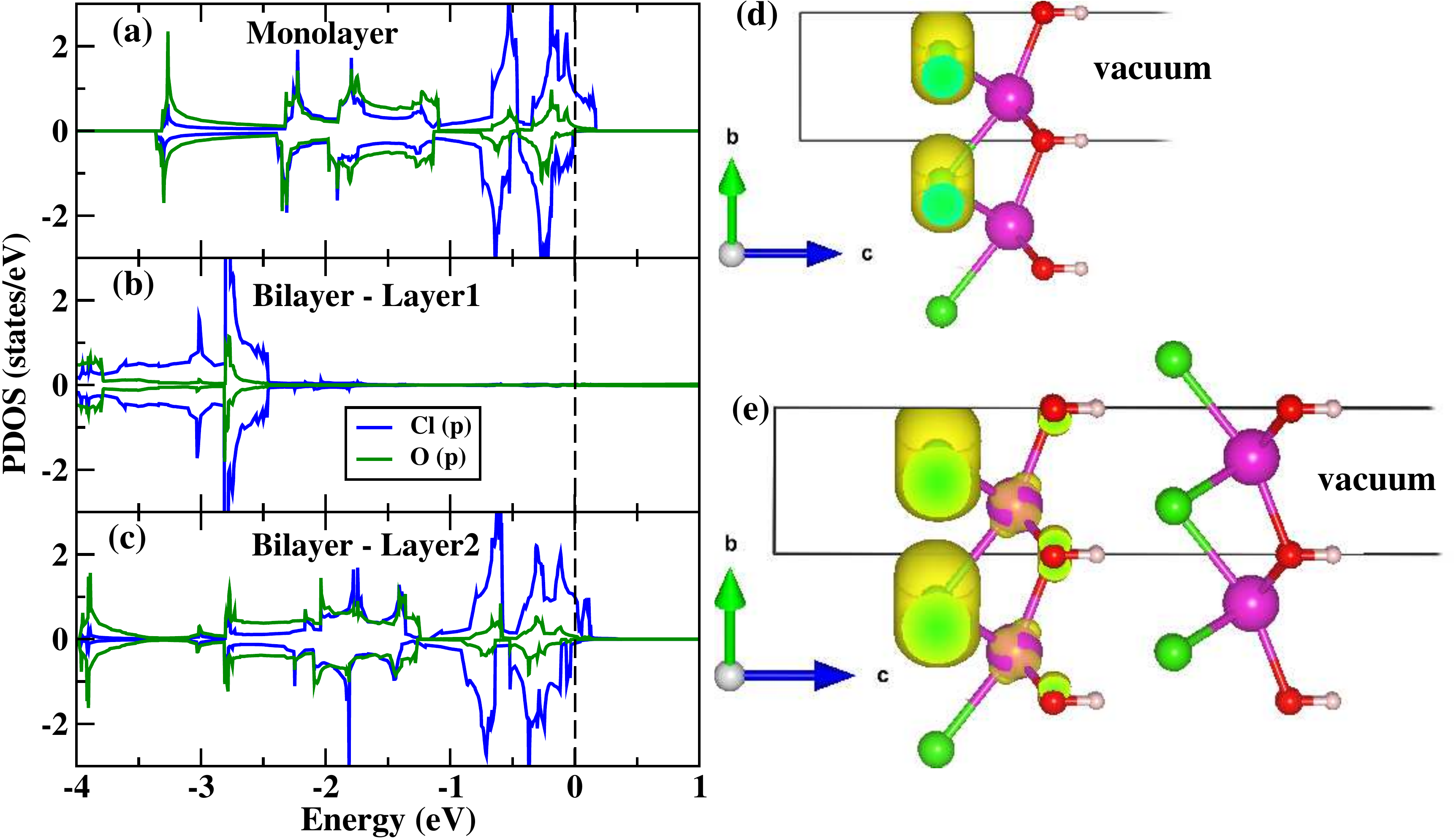}
	\caption{(Colour online) 
	Electronic Structure of monolayer and bilayer on application of electric field. Panel (a)-(c) shows projected DOS showing the band shifting induced by the application of an electric field E=0.6eV/$\AA$ on both (a) monolayer and (b)-(c) bilayer CdOHCl (panels (b) and (c) show the two different layers in the bilayer). The corresponding spin densities for monolayer (d) and bilayer (e) show the spatial distribution of magnetisation on specific layers.} 
\end{figure}


The full trend of magnetic moments versus magnitude of applied electric fields is shown in Figure 5(a). 
Even though the magnetic moments are rather small, this is to be expected: at variance with the case of homogeneous hole doping discussed in previous sections, here, by application of an electric field, one rather achieves doping (and related magnetic ground state) by shifting the relative energy position of the orbitals.
Upon varying the magnitude of the E-field in the monolayer, we find  that the smallest  ``critical" electric field to induce a magnetic moment on the $p$ bands amounts to 0.5 V/\AA. The latter is considered a value that can be \sout{easily} achieved by experimental standards. 
Upon reversing the direction of the electric field ({\em i.e.} applied in the -$z$ direction) up to a  magnitude  of -0.6 eV/\AA, no magnetisation is observed in the monolayer case, as  seen from Figure 5(a). Our estimates indicate that electric fields larger than 0.6 eV/\AA $\:$ are likely not easily accessible in experiments and hence  such very large negative electric fields are not further considered as source of induced magnetism in CdOHCl monolayers.

We interpret the asymmetric behaviour for positive and negative electric fields as related to the inherent polarity of the system. Indeed, Figure 5 (a) shows that there exists a certain direction of electric field  for which the activation of the Stoner mechanism is more favorable. 
This can be rephrased in terms of electric--field--induced shift of specific orbitals that mainly contribute to the Stoner instability: depending on the relative alignement of the electric field with respect to the intrinsic polarization due to the CdOHCl atomic stack, the tendency towards magnetism might be either favored or prevented, due to different orbital energy shifts for different signs of the applied field.  As such, a larger value of the external electric field is needed for the Stoner mechanism to be activated, as confirmed by Figure 5 (a). 

\subsection{Electric field applied to bilayer}

Interestingly, in the case of  bilayer of CdOHCl, the value of critical electric field  reduces drastically to 0.2 V/\AA, with respect to the monolayer situation. We plot in Figure 4 {\bf (b)-(c)} the layer projected DOS to demonstrate the DOS for the two layers to be rigidly shifted to very different energy regions. As such, the spin--polarization basically develops only on layer 2, as shown by the magnetization density plot reported in Figure 4 c). Our notation in Figure 4 is such that the layer labelled as ``2'' corresponds to the bottom layer in the spin density, showing a large (small) magnetisation on the Cl (O) atom, whereas the layer labelled as ``1'' corresponds to the top layer, which shows negligible magnetisation and, as seen in the DOS, is completely non magnetic and insulating. 
This asymmetric behaviour can be again interpreted along the line already mentioned:  the inherent polarisation present in the monolayer ({\em i.e.} related to the polar atomic layer stacking and related charge rearrangement) becomes further enhanced when considering the bilayer.
To further support this argument,   we plot in Figure  5 {\bf (b)-(d) } the averaged local potential in the bilayer for the cases of zero field and for finite values of electric field with opposite direction.  Let us \sout{start} first discuss the case of E=0 (cfr Figure 5 b) and take the Cl layers as reference. Even in the E=0 case, the environment of the two layers of Cl atoms are very different: one of the Cl atom ({\em i.e.} Cl2, as denoted in Figure 5 b) faces vacuum potential on one side and the potential of the Cd atom on the other side, while the  Cl atom on the other layer ({\em i.e.} Cl1, as denoted in Figure 5 b) faces the potential of the OH moiety on one side and that of the Cd atom on the other side [cfr atomic stacking previously shown in Figure 1 (a)]. The considerable asymmetry of the local potential of Cl1 and Cl2 (marked with a dashed arrow in the figure), already in the zero field case,  points to a sizeable electric polarisation in each layer; the latter is however not large enough for a magnetic moment to emerge. Nevertheless, a spin--polarization develops very easily in the presence of a small field of $0.2$V/\AA (cfr Figure 4 (b),(c),(e)), with the potential in case of a positive field applied along the +$z$ direction shown in Figure 5 (c). 

\begin{figure}[h]
	\centering
	\includegraphics[width=\columnwidth]{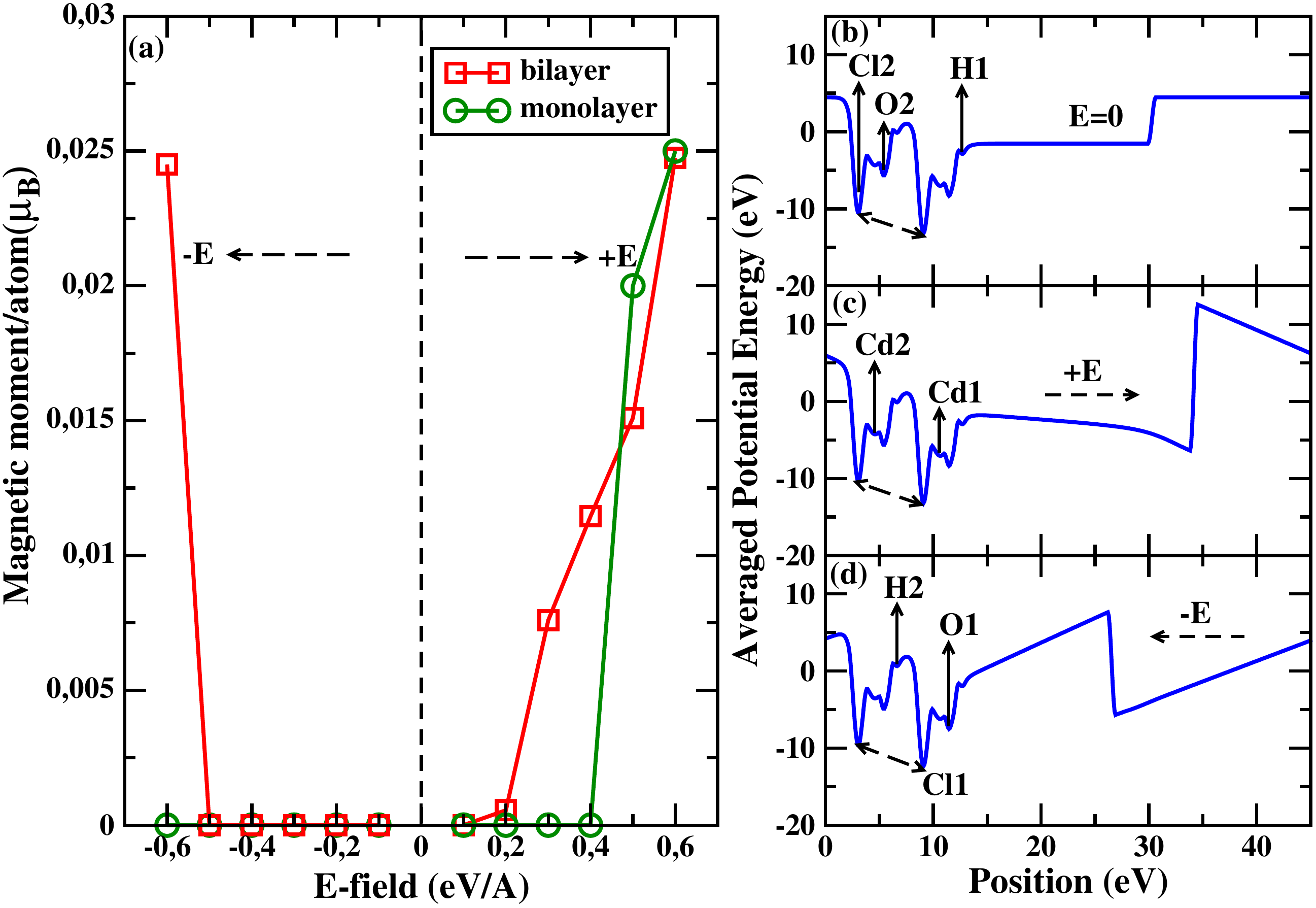}
	\caption{(Colour online) Panel (a) shows the comparison of magnetic moments upon application of the same electric field on both monolayer and bilayer samples of CdOHCl. Panels (b)-(d) of the figure shows local potential profiles in the CdOHCl bilayer. Panel (b) shows the E=0 case. Panel (c) shows the case of a E field of +0.6 eV/\AA in the +z direction while  panel (d) shows the potential in case of an  E field of -0.6 eV/\AA in the -z direction.} 
\end{figure}

When reversing the electric field in the bilayer, a much larger field ({\em i.e.} about   0.6 V/\AA) is needed to drive a magnetic solution, as shown in Figure 5(a). 
Correspondingly, the effect on the local potential upon reversing the field can be seen in Figure 5 (d). 
We infer that a  much larger value of electric field is  required to activate the Stoner mechanism and stabilise a magnetic state in the case of negative field, due to the previously discussed asymmetry; still, the estimated value of electric field in the negative $z$ direction  falls within the range of experimentally plausible values, so that a magnetic solution can be still driven. 
Finally, we underline that doping the bilayer and driving it into a magnetic state by means of application of E field is far easier than achieving the same aim in the monolayer case, as shown by the lower value of critical field.  

\section{Conclusion}

We have put forward CdOHCl, a polar layered material, as a candidate to achieve unconventional magnetism via i) hole-doping in both bulk and atomically-thin films and ii) electrostatic doping in monolayer and bilayers. By means of first-principles simulations, we have  shown that the bulk system is prone to a Stoner instability as the Fermi level is tuned by hole doping across such DOS singularities. Indeed, at reasonable values of doping fraction, a half-metallic ferromagnetic phase is  stabilized, magnetism primarily arising from $p$ orbitals of Cl and O, with large saturation magnetisation close to 1$\mu_B$ per hole. Being a van der Waals material, CdOHCl should be potentially exfoliable into few-layers or even monolayer atomically thin films, as suggested by our calculated exfoliation energies. We find that homogeneous hole doping is also effective in pushing the CdOHCl monolayer close to a Stoner instability, with the half-metallic ferromagnetism developing at carrier densities of the order of 10$^{14}$ cm$^{-2}$. 
Finally, the possibility of electrostatic doping has been addressed by applying a perpendicular electric field on both monolayer and bilayer phases of CdOHCl. The doping-induced magnetic transition is predicted to occur at experimentally accessible electric-field strengths of  $\sim$ 0.2 and 0.5 V/\AA~ for the bilayer and monolayer, respectively, where the reduced strength of the electric field required for the bilayer is attributed to the intrinsic polarity of the system. We hope our study will open up possibilities of observing such effects in experimental studies and of designing potential spintronic devices, exploiting the appealing half-metallic feature.

\section{Acknowledgements} HB thanks The Abdus Salam International Centre for Theoretical Physics (ICTP) for funding through the Training and Research in Italian Laboratories Programme (TRIL) in collaboration with Consiglio Nazionale delle Ricerche (CNR) Instituto SPIN. HB also thanks CINECA for allocation of a Class C project No. IsC66\_I-2DFM through the Italian Supercomputing Resource Allocation (ISCRA).  The calculations were performed using the resources of Gabriele d'Annunzio Universita degli Studi di Chieti and CINECA Supercomputing Centre. HB  is currently funded by the Austrian Science Foundation FWF through project TOPOMAT. P.B and S.P. acknowledge financial support from the Italian Ministry for Research and Education through PRIN-2017 projects
“Tuning and understanding Quantum phases in 2D materials - Quantum 2D” (IT-MIUR Grant No. 2017Z8TS5B) and “TWEET: Towards Ferroelectricity in two dimensions” (IT-MIUR Grant No. 2017YCTB59), respectively.

\section*{References}
\bibliography{ref}
\bibliographystyle{iopart-num}

\end{document}